\documentclass{PoS}
\usepackage{amsmath,amssymb,amsfonts}
\usepackage{slashed}
\usepackage{bbm}
\usepackage{cite}
\usepackage[utf8]{inputenc}
\usepackage[english]{babel}

\title{MCRG Flow for the nonlinear Sigma Model}
\ShortTitle{MCRG Flow for the nonlinear Sigma Model}
\author{Bjoern H. Wellegehausen\\%
        Justus-Liebig-University Giessen\\
        E-mail: \email{bjoern.wellegehausen@theo.physik.uni-giessen.de}}
\author{\speaker{Daniel Koerner} and Andreas Wipf\\%
        Friedrich-Schiller-University Jena\\
        E-mail: \email{daniel.koerner@uni-jena.de}, \email{wipf@tpi.uni-jena.de}}

\abstract{A study of the renormalization group flow in the three-dimensional
nonlinear O(N) sigma model using Monte Carlo Renormalization
Group (MCRG) techniques is presented. To achieve this, we combine
an improved blockspin transformation with the canonical demon method to
determine the flow diagram for a number of different truncations.
Systematic errors of the approach are highlighted. Results are discussed 
with hindsight on the fixed point structure of the model and
the corresponding critical exponents. Special emphasis is drawn on
the existence of a nontrivial ultraviolet fixed point as required for 
theories modeling the asymptotic safety scenario of
quantum gravity.}

\FullConference{31st International Symposium on Lattice Field Theory LATTICE 2013\\
		 July 29 - August 3, 2013\\
		 Mainz, Germany}

\graphicspath{{./img/}{./fig/}}

\newcommand{\vPhi}{\vec{\Phi}}
\newcommand{\vphi}{\vec{\phi}}

\begin{document}

\section{Asymptotic Safety}
\noindent In a renormalization group approach to QFT the theory 
at momentum scale $k$ is described by the effective average action
 $\Gamma_{k}$. It is obtained by choosing a microscopic action at
some cutoff-scale $\Lambda$ where $\Gamma_{\Lambda}=S_{mic}$
and integrating out fluctuations with scales between $\Lambda$ and
$k\ll\Lambda$ using the renormalization group flow. We call a theory 
\emph{fundamental} if
it is valid on all scales, i.e. the limit $\Lambda\rightarrow\infty$ 
and $k\rightarrow 0$ exists when one fine-tunes only a small number
of parameters. For gravity, we know the infrared
theory very well - it is given by the Einstein-Hilbert action. However, the
perturbative approach to find a formulation that is valid at small 
distances and high momenta reveals severe divergences: gravity
is not renormalizable in a perturbative manner. But it should still be 
possible to renormalize gravity nonperturbatively at a non-Gaussian 
ultraviolet fixed point with a finite
number of relevant directions (\emph{asymptotic safety scenario} 
\cite{weinberg80}). 
In this contribution, we study the nonlinear O(N)-models in $D=3$ 
dimensions, which are expected to show a nontrivial UV fixed point
\cite{percacci08,flore12}. The action is
\begin{equation}
S = \frac{1}{2g^{2}} \int 
d^{3}x\;\partial_{\mu}\vec{\phi}\partial^{\mu}\vec{\phi}\text{, \; with 
}\vphi\in\mathbb{R}^{N},\quad\vphi\vphi=1 .
\label{eq:action}
\end{equation}
Our goal is to obtain the global flow diagram from lattice simulations and 
determine its fixed point structure. To achieve this, we utilize 
Monte Carlo Renormalization Group (MCRG) techniques.

\section{MCRG method}

\begin{figure}[h]
\centering
\includegraphics[width=250px]{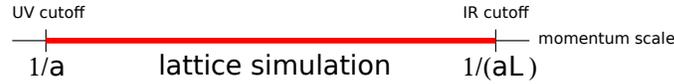}
\caption{A lattice simulation integrates out all the momenta between
the upper cutoff $\Lambda=a^{-1}$ and the lower cutoff $(aL)^{-1}$.}
\label{fig:eff_action2}
\end{figure}

\begin{figure}[h]
\centering
\includegraphics[width=250px]{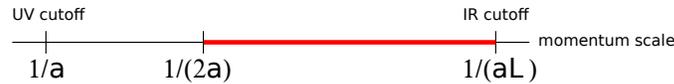}
\caption{A blockspin transformation with scale factor $2$ reduces the 
upper cutoff $\Lambda\rightarrow\Lambda'=\Lambda/2$, 
yet leaves the lower cutoff $(aL)^{-1}$ unchanged.}
\label{fig:eff_action3}
\end{figure}

\noindent On the lattice, the discrete momenta are cut off by the 
inverse lattice spacing $a^{-1}$ in the UV and the inverse box 
length $(aL)^{-1}$ in the IR and a lattice simulation is equivalent 
to integrating out all fluctuations in between (see figure \ref{fig:eff_action2}). 
The correlation 
functions of the theory are determined by direct measurement of 
lattice operators. By applying a blockspin transformation with scale 
factor $b$, the upper cutoff is reduced whereas the lower cutoff 
does not change: 
$a^{-1}\rightarrow (ba)^{-1}$, $(aL)^{-1}\rightarrow (baL/b)^{-1}=(aL)^{-1}$,
see figure \ref{fig:eff_action3}.
At the reduced cutoff, we define an \emph{effective} theory such that the IR
physics of the original and effective theory coincide, i.e. both 
theories lie on the same RG trajectory. In the RG picture, we have 
obtained the effective theory by integrating out the momenta 
from $a^{-1}$ until $(ba)^{-1}$. \\ 
In our setup, we use a local HMC algorithm for the O(N)-valued field
to compute a Markov-chain of configurations. This algorithm allows
to simulate arbitrary N with high acceptance rate. Also, it is
straightforward to add further operators 
to the action,
\begin{equation}
S[\vphi]=\sum_i g_i S_i[\vphi].
\label{eq:action}
\end{equation}
On each configuration, we apply a 
blockspin transformation to integrate out fluctuations and on the
blocked configurations, we use the canonical demon method 
\cite{hasenbusch95} to determine the couplings $g_{i}^{blocked}$ 
of the effective theory. In this way, we have access to the discrete 
beta function 
$\tilde{\beta}_{i}=\tilde{\beta}(g_{i}) = g_{i}^{blocked}-g_{i}$
and the discrete stability matrix
 $S_{ij}=\frac{\partial \tilde{\beta}_{i}}{\partial g_{j}}$. The 
eigenvalues of the stability matrix in the vicinity of a fixed
point can be related to its critical exponents.

\section{Systematic errors}

\noindent After comparing RG trajectories from different lattice sizes, no
finite volume effects are visible in our results for the 
$32^3\rightarrow 16^3$ simulations. Also, we expect discretisation
errors to be small near the critical line, where the correlation length
diverges in the continuum limit. However, using an effective 
action in the demon method inevitably leads to truncation errors.
In particular, the semi-group property of the combined transformation 
$R_{b}$ of blockspin transformation and demon method is violated:
$R_{b}\circ R_{b} \neq R_{b^2}$. Concerning the RG trajectories, this
leads to a discrepancy in the effective couplings of a chain of two transformations, 
with scale factor $b$ each, compared to a single transformation with scale
factor $b^2$.  In order to minimize this discrepancy we add further 
operators to the effective action \ref{eq:action} following a 
derivative expansion. Our best truncation consists of four operators 
which represent all possible operators of this model up to 
fourth order in the momenta:
\begin{align}
	S_{0} &= -\int d^{3}x\; \vphi\partial_{\mu}\partial^{\mu}\vphi, \\
	S_{1} &= -\int d^{3}x\; \vphi(\partial_{\mu}\partial^{\mu})^2\vphi, \\
	S_{2} &= -\int d^{3}x\; (\vphi\partial_{\mu}\partial^{\mu}\vphi)^{2}, \\
	S_{3} &= -\int d^{3}x\;
	(\vphi\partial_{\mu}\partial^{\nu}\vphi)(\vphi\partial^{\mu}\partial_{\nu}\vphi).
\end{align}
A second approach to overcome this problem is to use an improved
blockspin transformation\cite{hasenfratz84},
\begin{equation}
	\vPhi_{\tilde{x}} \propto P\left( \text{exp}\Big(C\; \vPhi_{\tilde{x}}\;\sum_{x\in\Lambda_{\tilde{x}}}\vphi_{x}\Big)\right),
\label{eq:blockspintrafo}
\end{equation}
where the blocked spin $\vPhi_{\tilde{x}}$ is drawn from a probability
distribution that takes into account a local neighbourhood 
$\Lambda_{\tilde{x}}$ of the
original spins: $\sum_{x\in\Lambda_{\tilde{x}}}\phi_{x}$. $C$ is
a temperature-like parameter that allows to introduce additional
noise. We parametrize $C=\sum_{i} c_{i}g_{i}$ such that for 
$g\rightarrow\infty$, $C\rightarrow\infty$ holds and no additional
noise is introduced in the far broken regime, where the spins are
aligned uniformly. The parameters $c_{i}$ are chosen such that 
truncation errors are minimal. To that end, we directly simulate
the ensemble of the truncated action 
and determine its correlation functions. Since blockspin
transformations do not change the IR physics, the systematic
effect of truncation will be visible as differences in the
correlation functions of the original and the truncated ensemble.
\begin{figure}
\centering
\input{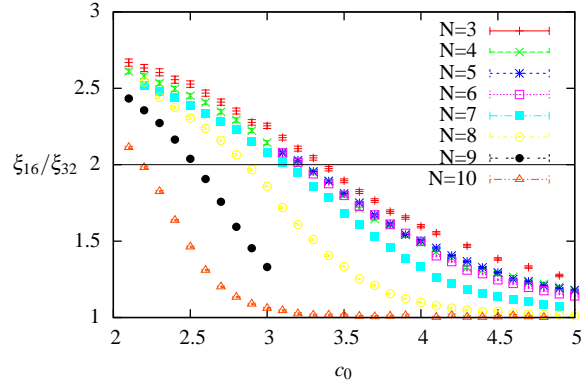}
\caption{The ratio of correlation lengths $\xi_{32}$ in the original
ensemble and $\xi_{16}$ in the truncated ensemble is used to fix
the optimization constant $c_0$ for different $N$. A value of
$\xi_{16}/\xi_{32}=2$ is expected to minimize the truncation errors.}
\label{fig:optim}
\end{figure}
An optimization of the blockspin transformation is equivalent 
to a minimization of the difference between the correlation functions. 
In general,
the optimal value depends on the coupling constants, lattice size,
target manifold and number of RG steps. As an 
approximation, we only consider the correlation length, which
increases by a factor $b$ after blocking: $\xi'=b\xi$. Figure 
\ref{fig:optim} shows that it is indeed possible to find an optimal
value $C=c_0g^*$ for the blockspin transformation (\ref{eq:blockspintrafo}) 
and a simple
one-parameter effective action at the fixed point coupling $g^*$.
From an RG
point of view, we use the fact that the location of the
renormalized trajectory, which connects the fixed points of the
RG flow, depends on the RG scheme. The optimal scheme 
causes the renormalized trajectory to lie closest to a given
truncation.

\section{Flow diagram}

\begin{figure}
\centering
\input{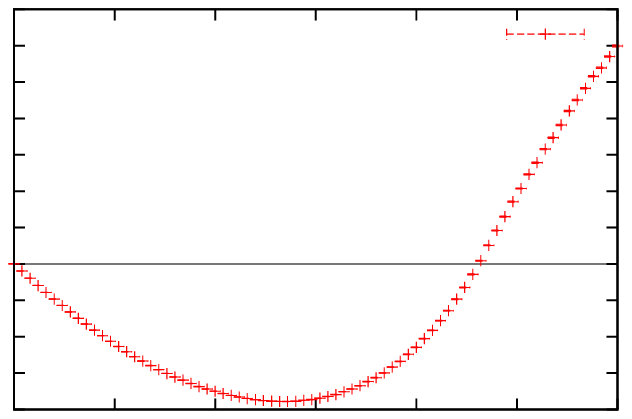}\hspace{20px}
\input{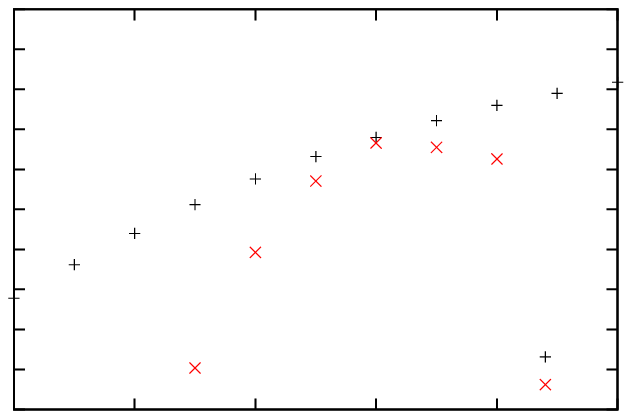}
\caption{\textbf{Left panel:} the lattice beta function shows a 
Gaussian and a non-Gaussian fixed point with UV-attractive direction. 
\textbf{Right panel:} the estimate for the critical exponent of 
the correlation length $\nu$ seems to improve for larger $N\leq6$,
but then deviates again.}
\label{fig:1param}
\end{figure}

\begin{figure}
\centering
\input{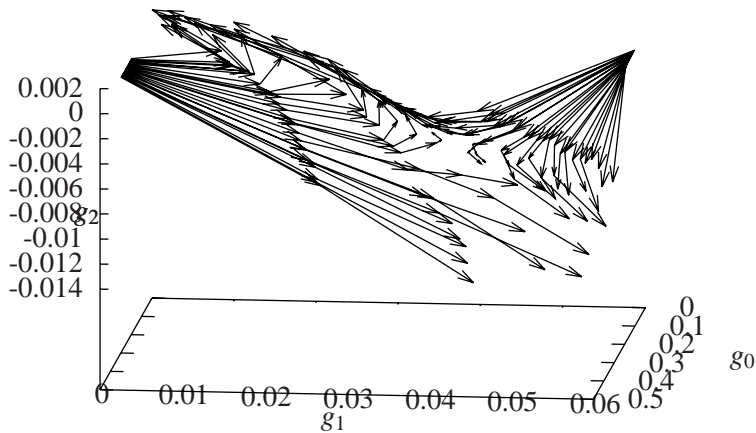}
\caption{Using a \emph{shooting} technique, we show trajectories
of the RG flow for the three-parameter effective action that flow
towards the renormalized trajectory in the vicinity of the 
non-Gaussian fixed point.}
\label{fig:3param}
\end{figure}

\noindent The beta function for the simplest possible truncation,
consisting of only a nearest-neighbor operator, shows a Gaussian 
fixed point at zero coupling and a non-Gaussian fixed point with 
a UV-attractive direction (see Fig. \ref{fig:1param}, left panel). 
From the
slope of the beta function, we can already determine the critical exponent $\nu$ 
of the correlation length (see Fig. \ref{fig:1param}, right 
panel) and we see that our result provides a reasonable estimate only 
for $N=6$. For values $N\neq6$, our estimate deviates from the
comparable result in the literature \cite{antonenko98}.
\begin{figure}
\centering
\input{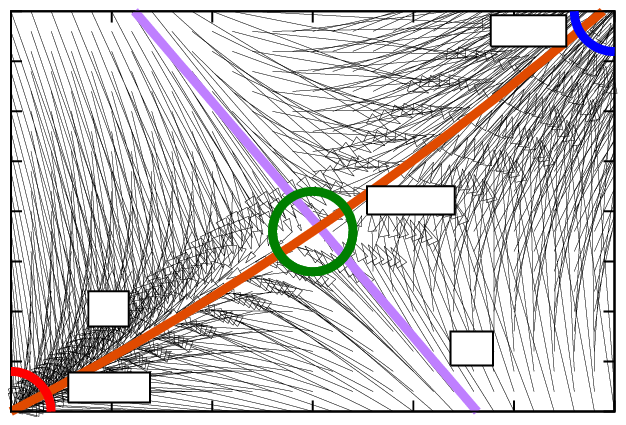}\hspace{20px}
\input{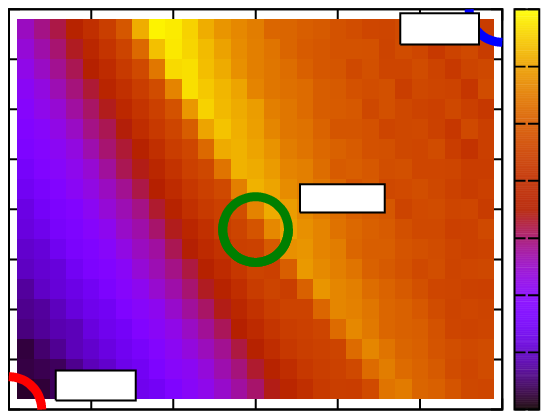}
\caption{\textbf{Left panel:} trajectories of the RG flow using the
two-parameter effective action. Abbreviations are given in the text. 
\textbf{Right panel:} eigenvalue of the stability matrix corresponding 
to the relevant direction. In the vicinity of a fixed point the 
eigenvalue is equal to the inverse critical exponent $\nu$.}
\label{fig:2param}
\end{figure}
In Fig. \ref{fig:2param} (left panel), we extended our truncation and
included a 
next-to-nearest-neighbour operator. We locate three fixed points
of the RG flow: a high temperature or Gaussian fixed point at
vanishing coupling (\emph{HT FP}), a low temperature fixed point at infinite
coupling (\emph{LT FP}) and a \emph{non-Gaussian fixed point} 
(\emph{NG FP}) with one IR-relevant and one IR-irrelevant direction.
Furthermore, we can clearly localize the critical line (\emph{CL})
which separates the flow diagram into the lower left part with 
trajectories flowing into the high-temperature fixed
point and the upper right part with trajectories flowing into the 
low-temperature fixed point. The critical trajectory itself flows
into the non-Gaussian fixed point. These three fixed points are
connected by the renormalized trajectory ($\emph{RT}$) and act
as infrared fixed points for the Heisenberg ferromagnet, which
"lives" on the $g_0$ axis. From the universality hypothesis, we
expect that the non-Gaussian fixed point corresponds to the
well-known Wilson-Fisher fixed point of the linear Sigma Model
and the structure found in \cite{bohr00} indeed matches our 
findings. However, the Heisenberg ferromagnet is an effective 
theory that is only valid at a fixed ultraviolet cutoff.
To obtain a fundamental theory that is complete both in the IR as well as
in the UV, we have to follow the renormalized trajectory,
where the non-Gaussian fixed point acts as an \emph{ultraviolet
fixed point} and, depending on the \emph{only} relevant direction, drives
the RG flow towards the low- or high-temperature fixed point.
Therefore, we clearly see the asymptotic safety scenario
fulfilled in this truncation. By computing the eigenvalues of
the stability matrix (see Fig. \ref{fig:2param}, right panel), 
we can extract a preliminary value for
the critical exponent at the non-Gaussian fixed point for $N=3$ of 
$\nu=0.64$, which is a significant improvement compared
to the 1-parameter effective action ($\nu=0.5$). Still, we
are about $10\%$ off from the predictions from a direct
computation of the thermodynamical critical exponents. At the trivial
fixed points, the critical exponent takes its trivial value $\nu=1$
or $\nu=-1$ as expected. These results are still preliminary and we 
are currently computing critical exponents for different values of $N$ and
will report our findings in a later publication \cite{mcrgpaper}. However, 
it is clear
that the presented method does not compete with traditional
Monte-Carlo methods in terms of precision but aims at a measurement
of the global flow diagram of the theory.\\
We continue
by computing trajectories for the 3-parameter truncation
and observe that only an irrelevant coupling is added to
the non-Gaussian fixed point. Figure \ref{fig:3param} 
shows trajectories that initially start in
the $g_2=0$ plane and descend below this plane towards the non-Gaussian
fixed point. Depending on the relevant direction, they
continue to flow along the renormalized trajectory towards 
the low-temperature or high-temperature fixed point.

\section{Conclusion}

\noindent We have shown that a combination of blockspin transformations
and demon method is suitable to obtain the global flow diagram of
the nonlinear Sigma Model in three dimensions. In contrast to the
traditional lattice matching technique, our analysis rests on single
trajectories rather than on long chains of trajectories. We therefore do not
need exponentially large lattices and a scan of the flow diagram
can be parallelized easily. By employing improved
blockspin transformations with a suitable optimization scheme, we
reduce the systematic errors that stem from a truncation of the effective
action.\\
 The flow diagram reveals two trivial IR fixed points that
correspond to absolute order and absolute disorder, respectively, and 
a nontrivial ultraviolet fixed point. This fixed point structure 
is stable against a change of truncation. In particular, we always
find only a single relevant direction for the UV fixed point.
Therefore, we conclude that the asymptotic safety scenario is 
realized for the nonlinear Sigma Model.\\
When we have finished our detailed simulations to obtain other critical 
exponents, we can compare our findings to large $N$ 
and FRG results. An interesting question would be to repeat the
present analysis in four dimensions, where the model is suspected
to be trivial. Furthermore, one may be interested to extend these
methods to fermionic models, e.g. the Thirring model which shows 
a rich flow diagram \cite{gies10}, or lattice quantum gravity \cite{Ambjorn:2001cv}.

\acknowledgments
\noindent We thank Raphael Flore for his active collaboration and 
Omar Zanusso and Luca Zambelli for helpful discussions. Simulations were
performed on the LOEWE-CSC at the University of Frankfurt and on the
Omega HPC cluster at the University of Jena.
	

\end{document}